\documentclass{aa} 

\usepackage{graphicx} 
\usepackage{txfonts} 
\usepackage{natbib}

\newcommand{\Mpc}{$h^{-1}$\thinspace Mpc}

\newcommand{\be}{\begin{equation}}
\newcommand{\ee}{\end{equation}}

\def\apj{ApJ}
\def\apjl{ApJL}
\def\apjs{ApJS}
\def\aj{AJ}
\edef\aap{A\&A}
\def\mnras{MNRAS}

\begin{document}   
\sloppy

\title{Towards understanding the structure of voids in the cosmic web} 

\author{ J. Einasto\inst{1,2,3} \and I. Suhhonenko\inst{1} \and
  G. H\"utsi\inst{1} \and E. Saar\inst{1,2} \and M. Einasto\inst{1} \and
  L. J. Liivam\"agi\inst{1} \and V. M\"uller\inst{4} \and
  A. A. Starobinsky\inst{5,6} \and E. Tago\inst{1} \and
  E. Tempel\inst{1}}

\institute{Tartu Observatory, EE-61602 T\~oravere, Estonia
\and 
Estonian Academy of Sciences,  EE-10130 Tallinn, Estonia
\and
ICRANet, Piazza della Repubblica 10, 65122 Pescara, Italy
\and
Leibniz-Institut f\"ur Astrophysik Potsdam, An der Sternwarte 16, D-14482 Potsdam,
  Germany
\and
Landau Institute for Theoretical Physics, RAS, 119334 Moscow, Russia
\and
Research Center for the Early Universe (RESCEU), Graduate School of Science,
The University of Tokyo,  113-0033 Tokyo, Japan
}

\date{ Received 12 May 2011/ Accepted 20 August 2011} 

\authorrunning{J. Einasto et al.}

\titlerunning{Void structure in the cosmic web} 

\offprints{J. Einasto, e-mail: einasto@aai.ee}
\abstract 
{According to the modern cosmological paradigm, cosmic voids form
  in low density regions  between filaments of galaxies and
  superclusters.} 
{ Our goal is to see how density waves of different scale combine
  to form  voids between galaxy systems of various scales.  }
{We perform numerical simulations of structure formation in cubes of
  size 100, and 256~\Mpc, with resolutions $256^3$ and $512^3$
  particles and cells. To understand the role of density perturbations of
  various scale, we cut power spectra on scales from 8 to 128~\Mpc,
  using otherwise in all cases identical initial random realisations. }
{ We find that small haloes and short filaments form all over the
  simulation box, if perturbations only on scales as large as 8~\Mpc\ are
  present.  We define density waves of scale $\geq 64$~\Mpc\ as large,
  waves of scale $\simeq 32$~\Mpc\ as medium scale, and waves of scale
  $\simeq 8$~\Mpc\ as small scale, within a factor of two. Voids form in
  regions where medium- and large-scale density perturbations combine
  in negative parts of the waves because of the synchronisation of phases of
  medium-  and large-scale density perturbations. In voids, the growth of
  potential haloes (formed in the absence of large-scale
  perturbations) is suppressed by the combined negative sections of
  medium- and large-scale density perturbations, so that their
  densities are less than the mean density, and thus during the
  evolution their densities do not increase. }
{ The phenomenon of large multi-scale voids in the cosmic web requires
  the presence of an extended spectrum of primordial density
  perturbations. The void phenomenon is due
  to the action of two processes: the synchronisation of density
  perturbations of medium and large scales, and the suppression of
  galaxy formation in low-density regions by the combined action of
  negative sections of medium- and large-scale density perturbations.}


\maketitle

\section{Introduction}
\label{sec:intro}

The goal of this series of papers is to study the role of
perturbations on various scales in  formation  the cosmic web.
\citet{Einasto:2011} used wavelet techniques to understand the
formation of rich systems of galaxies -- clusters and
superclusters. They conclude that superclusters are objects where
density waves of  medium and large scales combine {\em in similar
  phases to generate high density peaks}.  Similarly, voids are regions
in space where medium- and large-scale density perturbations combine in
similar {\em under-density phases} of waves. \citet{Suhhonenko:2011}
demonstrated that the properties of the cosmic web depend essentially on
density perturbations of small and medium scales, whereas perturbations
of  large scale $\geq 100$~\Mpc\ modulate the richness of galaxy
systems from clusters to superclusters, and make voids emptier.  This
paper is devoted to the study of the influence of medium- and large-scale
density waves on the structure and evolution of voids in the cosmic web.

The cosmic web was first openly discussed at the IAU Symposium on
Large Scale Structure of the Universe \citep{Longair:1978}.  At this
symposium, four groups reported results of studies of the three-dimensional
distribution of galaxies in space using  available data for the 
redshifts of galaxies. The presence of voids in the distribution of
galaxies was reported by \citet{Joeveer:1978a}, \citet{Tarenghi:1978}, 
\citet{Tifft:1978}, and \citet{Tully:1978} in the Perseus-Pisces, Hercules, 
Coma, and Local superclusters, respectively.

The main results reported in this symposium were that: (1) galaxies,
groups, and clusters of galaxies are not randomly distributed but form
chains, converging in superclusters; (2) the space between galaxy
chains contains almost no galaxies and forms voids of diameters $20
\dots 70$~\Mpc; (3) superclusters are not isolated systems, but are
connected by galaxy filaments to a connected network -- the
supercluster-void network \citep{Joeveer:1978,
  Einasto:1980,Zeldovich:1982, Oort:1983} or the cosmic web
\citep{Bond:1996}.  These early results were confirmed by the Second
Harvard Sky Survey \citep{de-Lapparent:1986,Geller:1989} and the
discovery of the very large Bootes void by
\citet{Kirshner:1981,Kirshner:1987}.

The presence of long essentially one-dimensional galaxy and
group/cluster chains, and the elongated shape along the chain of the
central cluster galaxies (often supergiant galaxies of type cD)
suggests that galaxies and groups/clusters of the chain had formed
within the chain simultaneously with the formation of the whole cosmic
web. This occurred in the gaseous phase of the structure evolution,
which allowed the dissipation and cancelling of velocities
perpendicular to the chain axis.  These data gave strong support to
the \citet{Zeldovich:1970,Zeldovich:1978} pancake scenario of galaxy
formation.  The observed distribution of galaxies was quite similar to
the distribution of simulation particles in a two-dimensional
numerical simulation of the evolution of the structure of the
Universe, prepared by Shandarin (1975, private communication) and
published by \citet{Doroshkevich:1980ij}. In this simulation a network
of high- and low-density regions was seen: high-density regions form
cells that surround large under-dense regions. Subsequent
three-dimensional simulations confirmed this picture
\citep{Klypin:1983,Melott:1983}.

However, some important differences between the model and observations
were evident.  First of all, there exists a rarefied population of
simulation particles in voids that is absent in real data. This was the first
indication  of physical biasing in galaxy formation
(the term ``biased galaxy formation'' was introduced a few years later
by \citet{Kaiser:1984}, see also \citet{Bardeen:1986}).  A theoretical
explanation of the absence of galaxies in voids was given by 
\citet{Einasto:1980} (see also a more detailed discussion by
\citet{Einasto:1994}).  A simple analytical solution of the
cosmological evolution of the density of matter shows that in
over-dense regions the density increases until the matter collapses to
form compact objects (pancaking by \citet{Zeldovich:1970}).  In
contrast, in under-dense regions the density decreases but never
reaches a zero value -- gravity cannot evacuate voids completely.

Even early studies proposed that the cosmic web has a hierarchical
structure.  The parts of the web formed by objects of different mass
or luminosity have different characteristic sizes.  As voids are
defined by the web, the void assembly has several important
properties: voids defined by more luminous (or massive) objects have
larger diameters, and voids defined by clusters/galaxies of certain
luminosity contain substructure formed by less luminous objects.

According to the present cosmological paradigm, all structural
elements of the Universe were formed by the growth of initial small
density perturbations created during the very early phase of the
evolution of the Universe. To these elements belong galaxies, groups,
clusters, and their systems, such as filaments and superclusters.
Rich superclusters form a cellular distribution, with large voids
surrounded by rich superclusters.  The characteristic diameter of
these supervoids is of the order of 100~\Mpc\ \citep{Joeveer:1978,
  Kirshner:1981, Einasto:1994wd, Einasto:1997lh}.  Supervoids are not
empty, but contain a hierarchy of voids \citep{Einasto:1989a,
  Martel:1990, van-de-Weygaert:1993, Lindner:1995ui, Muller:2000,
  Gottlober:2003, Aragon-Calvo:2007, von-Benda-Beckmann:2008qf,
  van-de-Weygaert:2009pd,van-de-Weygaert:2009, Aragon-Calvo:2010,
  Jones:2010cr}.

Among the early studies of the void evolution we mention
\citet{Hoffman:1982, Hoffman:1983}, \citet{Peebles:1982},
\citet{Icke:1984}, \citet{Fillmore:1984}, \citet{Bertschinger:1987},
among others.  The study of the hierarchical evolution of voids was
pioneered by \citet{Dubinski:1993}.  \citet{Sahni:1994} described the
evolving void hierarchy within the context of the adhesion theory.
\citet{Sheth:2004ly} developed the excursion set (extended
Press-Schechter) description of a void hierarchy in the dark matter
distribution, followed by \citet{Furlanetto:2006} comparing this with
the void hierarchy in galaxy populations.

Superclusters are connected by filaments of galaxies, i.e voids have
substructure.  Observationally, this was already evident in  early
 void studies \citep{Joeveer:1978a, de-Lapparent:1986}.
Theoretical discussion of the void substructure has been presented by
\citet{Regos:1989}, \citet{Martel:1990}, \citet{van-de-Weygaert:1993},
\citet{Goldberg:2004,Goldberg:2005}, and many others.  The skeleton of
the cosmic web was  discussed by \citet{Hahn:2007ve},
\citet{Forero-Romero:2009}, \citet{Sousbie:2008ul, Sousbie:2009lq},
\citet{Aragon-Calvo:2010wd}, \citet{Bond:2010rr, Bond:2010cr},
\citet{Shandarin:2010}, and \citet{Einasto:2011}.

It is generally accepted that the initial density perturbations had a
smooth, extended power spectrum (quasi-flat in terms of metric perturbations
$n_s\approx 1$) and a random (Gaussian) distribution
of perturbation phases.  The amplitude of perturbations ($\Delta
\propto k^3 P(k)$) is larger at short wavelengths and per $\delta \ln
k$, where $k$ is the wavenumber, and $P(k)$ is the power spectrum of
perturbations.  For this reason, small objects (mini-haloes in
numerical simulations and dwarf galaxies in the real Universe) should
form first.  The early formation of dwarf galaxies is confirmed by
observation of very distant galaxies \citep{Beckwith:2006}.  These
early galaxies grow by the attraction of more primordial matter and by
clustering, as originally suggested  by \citet{Peebles:1971}.

Initial small-scale perturbations were present everywhere, and this
raises the question: why do voids not contain galaxies, even dwarf
galaxies?  This question was asked more specifically by
\citet{Peebles:2001fk}.  The deepest voids in the simulated dark
matter distribution are never completely empty; they still contain
low-mass condensations of primordial matter -- mini-haloes. Therefore,
voids could be the environment in which faint dwarf galaxies are most
likely to reside. Why is this not the case?

This ``Peebles question'' has stimulated a number of studies, both
observational and theoretical, to find dwarf void galaxies and to
study either analytically or by numerical simulations void regions of
the Universe. Studies of void galaxies were made by
\citet{Szomoru:1996}, \citet{Grogin:1999}, \citet{Gottlober:2003},
\citet{Rojas:2004}, \citet{Hoeft:2006}, \citet{Stanonik:2009},
\citet{Kreckel:2011, Kreckel:2011a}, and others.  Among recent
large-scale surveys, we mention here \citet{Hoyle:2004uq},
\citet{Croton:2004vn}, \citet{Conroy:2005rt}, \citet{Patiri:2006dq},
and \citet{Tinker:2007ai}.  \citet{Karachentsev:2003c,
  Karachentsev:2004, Karachentsev:2007b} studied the Local Volume out
to a distance 10~\Mpc, where they found about 550 mainly very faint
galaxies. A void galaxy survey (VGS) was initiated by van de Weygaert
and collaborators in our local neighbourhood out to redshift $z=0.02$
\citep{Stanonik:2009, Kreckel:2011a, Kreckel:2011, 
  van-de-Weygaert:2011kl}.  These studies suggest that extremely faint
galaxies do not fill voids, but are located in the vicinity of
brighter galaxies.

To investigate the properties of galaxies in voids, the halo
occupation distribution model is used to determine the relationship
between galaxies and dark matter haloes (see \citet{Seljak:2000},
\citet{Cooray:2002zr}, \citet{Berlind:2002ly}, \citet{Zehavi:2004la},
\citet{Zehavi:2005dz}, \citet{Zheng:2007fu},
\citet{van-den-Bosch:2007hc},
\citet{Tinker:2006kl,Tinker:2007ai,Tinker:2008fj,Tinker:2008tg},
\citet{White:2007ij}, and \citet{Padmanabhan:2009bh}).  The results
obtained by analysing of observational data are in good agreement with
the results of semi-analytic models, hydrodynamic cosmological
simulations, and high-resolution collisionless simulations
\citep{Kravtsov:2004ve, Zheng:2005pb, Conroy:2006ys,
  von-Benda-Beckmann:2008qf, Tikhonov:2009nx}.  In particular, the
simulations of \citet{Tikhonov:2009nx} and \citet{Ishiyama:2011fk}
show the presence of very low-mass haloes in voids, but their mass is
probably too low for star formation to be possible.  The comparison of
the distribution of model haloes and dwarf galaxies shows that haloes
should have a circular velocity of at least $\sim 35$ km/s in order to
form a galaxy.  \citet{Tinker:2008tg} showed that the smallest void
haloes have so little mass that no galaxy formation is expected
according to the model.

The absence of dwarf galaxies in voids can also be explained by
gasodynamical processes.  Using high-resolution hydrodynamical
simulations, \citet{Hoeft:2006}, and \citet{Hoeft:2010kx} showed that
photoionisation by the UV radiation field is able to stop the cooling
and collapse of gas in dwarf galaxy haloes.  At the redshift $z=0$,
the characteristic mass scale of photo-evaporation corresponds to a
circular velocity $\sim 27$ km/s, in good agreement with other studies
cited above.

All studies suggest that galaxy formation is a threshold phenomenon.
In many of the studies cited above, it was demonstrated that galaxies
do not form in voids, when the void haloes have very low
masses. However, these studies did not explain {\em what is the
  physical reason for the difference between the typical halo masses
  in the void and supercluster environments?}  In this paper, we try
to find an explanation for the difference between the masses of haloes
in various global environments.

In the case of the presence of an extended primordial perturbation
spectrum, systems of galaxies are produced by an interplay of density
perturbations on all scales. It is clear that the difference between
the present distribution of galaxies and the very early distribution
of protogalaxies must have something to do with perturbations of a
typical scale-length that is larger than the scales responsible for
the formation of primeval small protohaloes.  This lead us to the
study of the influence of perturbations of various scale-lengths on
the evolution of the structure.  Preliminary results of this study
very interestingly demonstrated that the whole supercluster-void
network is caused by the interplay of medium- and large-scale
perturbations. These preliminary results were reported on several
conferences and summer-schools (see the web-site of Tartu Observatory
\footnote{\tt http://www.aai.ee/$\sim$einasto/reports.php} ).

In the present paper, we attempt to understand the influence of
perturbations of various scale on the evolution and structure of voids
in the cosmic web.  As in \citet{Suhhonenko:2011} and
\citet{Einasto:2011}, we use numerical simulations in boxes of various
scale-lengths from 100 to 256~\Mpc, calculated for power spectra cut
off above different scales from 8 to 128~\Mpc, to determine the
influence of perturbations of various scales on the formation and
internal structure of voids. We employ a wavelet technique to follow
the evolution of under-dense and over-dense regions in terms of
density waves of various scales.

The paper is composed as follows. In the next section, we describe
numerical models used in the study.  In Section 3, we perform a
wavelet analysis of our simulations.  In Section 4, we describe our
correlation analysis of wavelet-decomposed density fields. In Section
5, we investigate the structure of voids using haloes of various mass
as objects which define voids.  We discuss our results in Section 6,
and in our last Section we present our conclusions.

\section{Modelling the evolution of voids in the cosmic web}

Previous analyses of the observational galaxy samples and numerical
simulations have shown that in the formation of superclusters and
voids, large-scale perturbations play an important role.  Thus, to
understand the supercluster-void phenomenon correctly, we need to
perform numerical simulation in a box containing large waves.  On the
other hand, most systems of galaxies in the Universe are groups of
galaxies -- there are almost no very isolated galaxies far away from
groups.  The characteristic scale of groups is 1~\Mpc, thus the
simulation must have at least a resolution similar to this scale.

To have both a high spatial resolution and the presence of density
perturbations in a larger scale interval, we performed simulations in
boxes of sizes 100~\Mpc, and 256~\Mpc, and resolutions
$N_{\mathrm{grid}}^3 = 256^3$ and $N_{\mathrm{grid}}^3 = 512^3$.  The
main parameters of our series of models are given in Table~\ref{Tab1},
where $L$ is the cube size, $N_{\mathrm{part}}$ is the number of
particles and cells used in simulations, and $M_{\mathrm{part}}$ is
the mass of a particle in units of $10^9~M_\odot$. We assumed the
cosmological parameters
\citep{Seljak:2005dk,Tegmark:2004a,Tegmark:2006rw}
$\Omega_{\mathrm{m}} = 0.28$ for the matter density, $\Omega_{\Lambda}
= 0.72$ for the dark energy density (in units of the critical
cosmological density), and $\sigma_8 = 0.84$ for the initial amplitude
parameter.  We use the notation $h$ for the present-day dimensionless
Hubble parameter in units of 100 km s$^{-1}$ Mpc$^{-1}$; in
simulations we used a value of $h=1$.

As we are interested in the study of the role of perturbations on 
different scales to the evolution of voids, we used simulations with
the full power spectrum, as well as with a power spectrum truncated at
wave-numbers $k_{\mathrm{cut}}$, so that the amplitude of the power
spectrum on large scales is zero: $P(k) = 0$, if $k<
k_{\mathrm{cut}}$, wavelength $\lambda_{\mathrm{cut}} =
2\pi/k_{\mathrm{cut}}$.  The cut scale in \Mpc,
$\lambda_{\mathrm{cut}}= {2\pi/k_{\mathrm{cut}}}$, is given in 
Table~\ref{Tab1}. The amplitude of a spectrum was set to zero for $k<
k_{\mathrm{cut}}$ during the calculation of the initial density field,
keeping all simulation parameters fixed across the full set of
realisations.

{
\begin{table}[ht]
\caption{Parameters of models.}
\begin{tabular}{lrrcc} 
\hline 
\noalign{\smallskip}
Model  &$L$& $\lambda_{\mathrm{cut}}$&
$N_{\mathrm{part}}$&$M_{\mathrm{part}}$\\
\hline 
&(1)&(2)&(3)&(4)   \\
\noalign{\smallskip}
\hline 
\noalign{\smallskip}
\\
M256.256    & 256 & 256    & $256^3$ & 77.72 \\
M256.064    & 256 & 64  & $256^3$ & 77.72  \\
M256.032    & 256 & 32  & $256^3$ & 77.72  \\
M256.016    & 256 & 16  & $256^3$ &  77.72 \\
M256.008    & 256 & 8   & $256^3$ & 77.72  \\
\\
L256.256    & 256 & 256    & $512^3$ &  9.714 \\
L256.128    & 256 & 128    & $512^3$ &  9.714 \\
L256.064    & 256 & 64  & $512^3$ &  9.714  \\
L256.032    & 256 & 32  & $512^3$ &  9.714 \\
L256.016    & 256 & 16  & $512^3$ &  9.714 \\
L256.008    & 256 & 8   & $512^3$ &  9.714  \\
\\
L100.100    & 100 & 100  & $512^3$ & 0.5583 \\
L100.032    & 100 & 32  & $512^3$ & 0.5583  \\
L100.016    & 100 & 16  & $512^3$ &  0.5583 \\
L100.008    & 100 & 8   & $512^3$ & 0.5583 \\
\noalign{\smallskip}
\hline
\label{Tab1}                        
\end{tabular}
\tablefoot{
\\
\noindent column 1: $L$ - size of the simulation box in \Mpc;\\
\noindent column 2: $\lambda_{\mathrm{cut}}$ - cut scale in \Mpc;\\
\noindent column 3: Number of particles;\\
\noindent column 4: Particle mass in $10^9~M_\odot$.}
\end{table}
}

\begin{figure*}[ht]
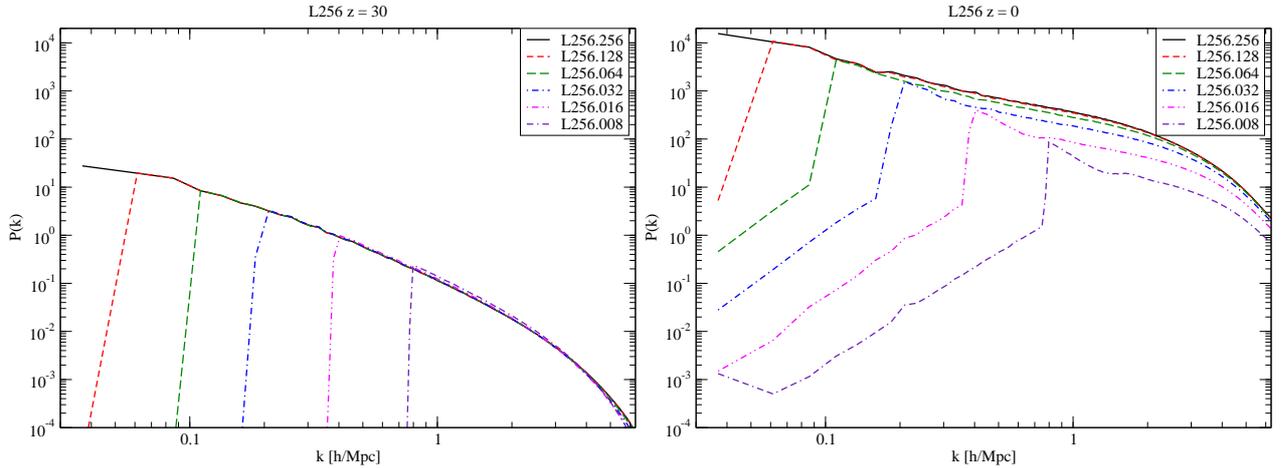
 
\centering 
\resizebox{0.45\textwidth}{!}{\includegraphics*{17248_fig1a.eps}}
\resizebox{0.45\textwidth}{!}{\includegraphics*{17248_fig1b.eps}}
\caption{The left and right panels show the power spectra for the
  models of the series L256 at the epochs $z = 30$ and $z = 0$,
  respectively. }
\label{fig:spec} 
\end{figure*}

As in \citet{Suhhonenko:2011}, we use the notations for our models
whereby the first characters M and L designate models with resolutions
of $N_{\mathrm{grid}} = 256$ and $N_{\mathrm{grid}} = 512$,
respectively. The subsequent number gives the size of the simulation
box, $L$, in \Mpc, and the next indicates the maximum wavelength used
in the simulation in \Mpc. The locations of the cells inside the
cubical density grid are marked by cell indices $(i,j,k)$.

For the models of the M256 series, we used in simulations the AMIGA
code \citep{Knebe:2001qa}.  This code uses an adaptive mesh technique
in the regions where the density exceeds a fixed threshold.  In this
code, gravity is automatically softened adaptively, so that the
softening length is near its optimum value in both high- and
low-density regions.  We chose a maximum level of eight refinements.
For models with a $512^3$ resolution, we used the GADGET-2 code with a
gravitational softening length of 10~$h^{-1}$\thinspace kpc (L100
models) and 20~$h^{-1}$\thinspace kps (L256 models)
\citep{Springel:2001,Springel:2005}. The simulation L100 was performed
at the Leibniz-Institut f\"ur Astrophysik Potsdam, and simulations
M256, and L256 at the High Performance Computing Centre of University
of Tartu.  The initial density fluctuation spectrum was generated
using the COSMICS code by \citet{Bertschinger:1995}\footnote{\tt
  http://arcturus.mit.edu/cosmics}; to generate the initial data, we
assumed the baryonic matter density $\Omega_{\mathrm{b}}= 0.044$.
Calculations started at an early epoch, $z=100$. For every particle,
we calculated the local density in units of the mean density, using
positions of 27 nearby particles.  All models of the same series have
the same realisation, so the role of different waves in models can be
easily compared.  To allow different models to be compared every
particle has an identification number, the same for all models of a
series. Particle positions and velocities were extracted for seven
epochs in the redshifts range $z=30, \dots, 0$.  Power spectra for the
models of the L256 series are shown in Fig.~\ref{fig:spec} for an
early epoch, $z = 30$, and for the present epoch $z = 0$.

For each particle, we also calculated the global density at the
location of the particle (for details see Appendix A.1). To find the
global density field, we used smoothing with the $B_3$ spline kernel
of scale $\sim 8$~\Mpc, which is rather close to smoothing with an
Epanechnikov kernel of the same scale. Smoothing of density fields
with different kernels is discussed by \citet{Martinez:2002ye}.  These
local and global density fields were calculated for all models and all
epochs, and were used in the subsequent ana\-lysis to select particles
belonging to a population with given properties, and to follow the
density evolution of the model. Hence, for each particle we stored
coordinates, local, and global density values. Particles were sorted,
thus their number in the file serves as an ID number.

To see the effects of density waves of different scale and to
understand the evolution of the density field, we use the wavelet
technique.  We use the \'a trous wavelet transform (for details, see
\citet{Martinez:2002ye} and Appendix A.2).  The field is decomposed
into several frequency bands as follows. The high-resolution (zero
level) density field was calculated with the $B_3$ spline kernel of
width equal to the size of one cell of the field, where every next
field was calculated with twice larger kernel.  Wavelets were found by
subtracting higher level density fields from the previous level
fields. In such a way, each wavelet band contains waves twice the size
of the previous band, in the range $\pm\sqrt{2}$ centered on the mean
(central) wave.  The sum of these bands restores the original density
field.  Using this technique, we calculated the density fields and
wavelets up to index 5 (6 for models of the M256 series).

\begin{figure*}[ht]
\centering
\resizebox{0.9\textwidth}{!}{\includegraphics*{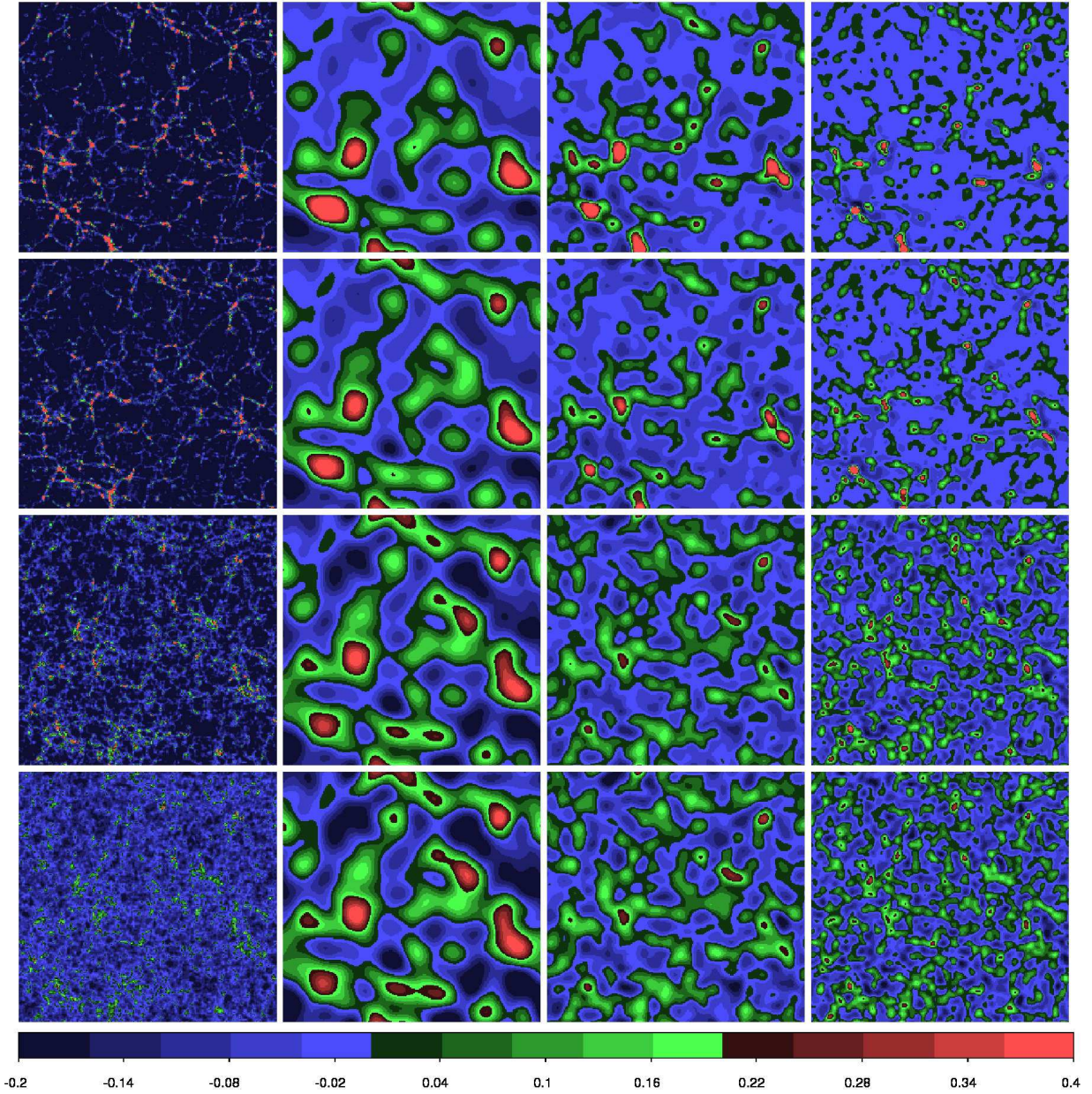}}
\caption{The high-resolution density fields of the model M256.256 are
  shown in the left column, at the $k=140$ coordinate. The second,
  third, and fourth columns show the wavelet $w5$, $w4$, and $w3$
  decompositions at the same $k$, respectively. The upper row gives
  data for the present epoch, $z=0$, the second row for the redshift
  $z=1$, the third row for the redshift $z=5$, and the last row for
  the redshift $z=10$. Densities are expressed on a linear scale.  In
  wavelet panels, green and red colours show the positive regions of
  wavelets, and the blue colour shows negative (under-density) wavelet
  regions. Colour codes are plotted at the bottom of the figure for
  wavelet $w3$ at epoch $z=10$.}
\label{fig:M256wav}
\end{figure*}

\begin{figure*}[ht]
\centering
\resizebox{0.9\textwidth}{!}{\includegraphics*{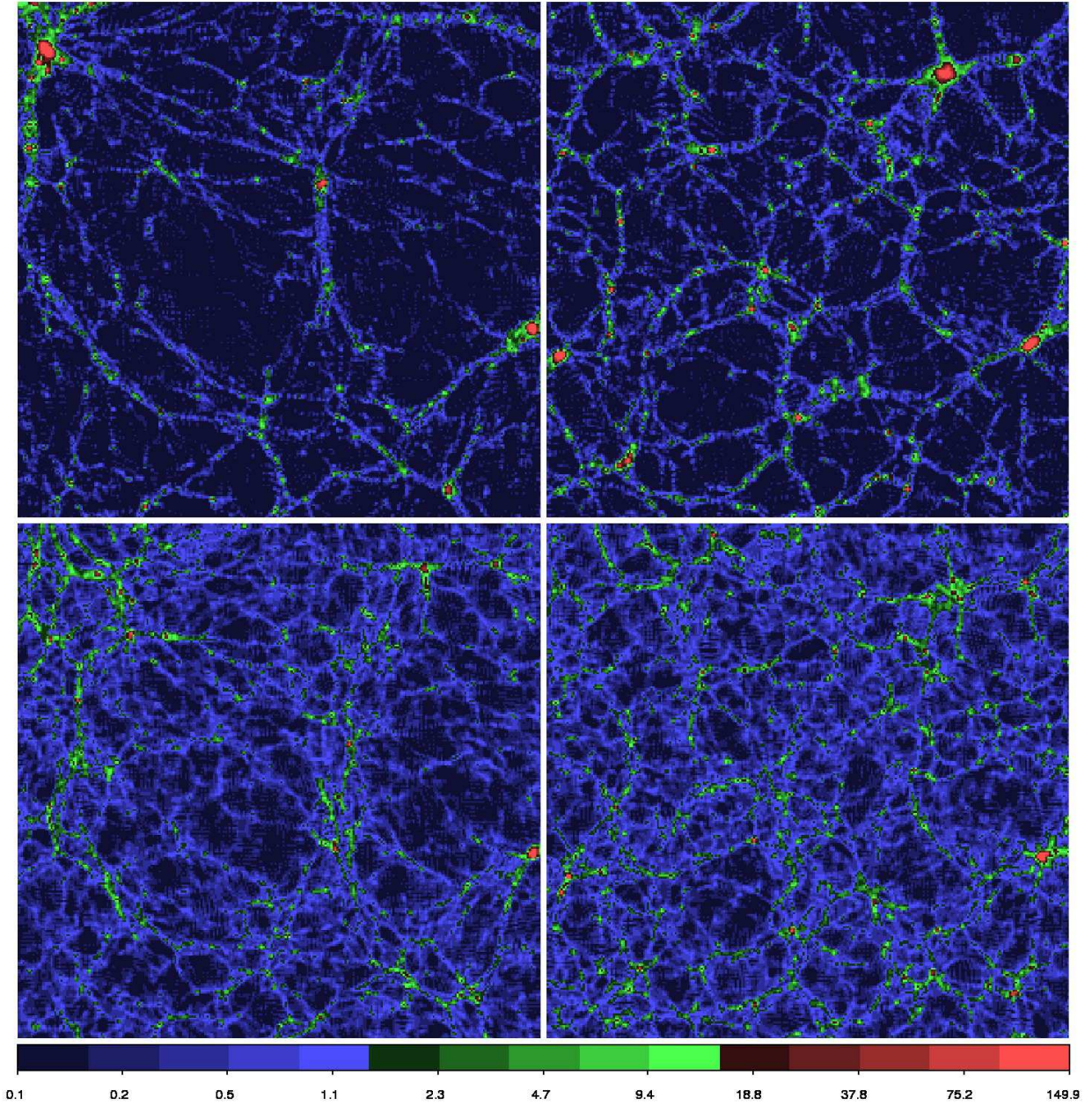}}
\caption{{Zoom-ins to the high-resolution density fields of the models
  L100.100 and L100.016, left and right columns, respectively.  Zoom
  factor is 2, central $50 \times 50$~\Mpc\ ($256 \times 256$ pixels)
  of all models are shown.  Upper panels are for the present
  epoch $z=0$, lower panels for the epoch $z=2$. All panels are at the
  $k=51$ coordinate.  Cross sections (beams) at coordinates $j=222$
  and $k=51$ for both models are shown in Fig.~\ref{fig:L100beam} for
  three redshifts to see the evolution of the density field and its
  wavelets.  In the upper left corner of the Figure there is a rich
  supercluster in the model L100.100, absent in the model L100.016.
  Both models have at the right edge of the Figure a rich cluster.
  This cluster is well seen in the Fig.~\ref{fig:L100beam} at $i =
  390$ coordinate.  Densities are expressed in the logarithmic scale,
  identical lower and upper limits for plotting with the SAO DS9
  package are used.  The border between the light blue and the dark
  green colours corresponds to the critical density $D_{\mathrm{loc}}
  = 1.6$,  which separates low-density haloes and haloes collapsed
    during the Hubble time \citep{Kaiser:1984, Bardeen:1986}.  Note
    that in both models and simulation epochs the majority of
    filaments in voids have densities below the critical density.} }
\label{fig:L100den}
\end{figure*}

The high-resolution density fields of the model  M256.256 for
the epochs $z=0,~1,~5,~10$ are shown in Fig.~\ref{fig:M256wav}.  The
high-resolution density fields of models  L100.100 and L100.016
for redshifts $z=0$ and $z=2$ are shown in  Fig.~\ref{fig:L100den}.
The scale of the cosmic web is rather different in models of different
cutoff scale. The dependence of the scale of the web on the maximal
wavelength of density perturbations was investigated in detail by
\citet{Suhhonenko:2011}.

To investigate the spatial structure of both the cosmic web and voids,
we found haloes using the adaptive Amiga Halo Finder (AHF) code
developed by \citet{Knollmann:2009}, with the number of particles in a
halo $N_{\mathrm{p}} \geq 20$.  Haloes and their parameters (masses,
virial radii, positions, velocities etc.)  were found for all models
and simulation epochs.

\section{Analysis of models}

To understand the evolution of the density field, and to see the
effects of density waves of different scale, we analyse the density
fields and their wavelet decompositions for various simulation epochs
and cut off scales.  We focus on the evolution of the following
properties:
\begin{enumerate}
\item the global patterns of the density field, using
  wavelets of different levels;
\item the fine structure of the density field for
  perturbations of various scales;
\item density perturbations of various scales;
\item the density distribution in void regions.
\end{enumerate}

\subsection{The evolution of the global patterns of density fields}

To compare the patterns of density fields and their wavelet
decompositions for various cosmic epochs, we use the model M256.256.

In Fig.~\ref{fig:M256wav}, we plot the high-resolution density field and
wavelets of the model M256.256 at the four redshifts: $z =
0,~1,~5,~10$. In the second column of Fig.~\ref{fig:M256wav}, the
wavelets of  order five are shown.  The characteristic scale of
  density perturbations for this wavelet is 64~\Mpc.  The upper
density levels used in plotting for $w5$ are 1.2, 0.7, 0.2, and 0.1,
for redshifts 0, ~1, ~5, and ~10, respectively. Lower limits to these
redshifts are $-0.6$,~$-0.35$,~$-0.1$, and $-0.05$.  In wavelets of the
order 4 and 3, a similar choice of colour limits is applied.  This
colour-coding of wavelets at different redshifts is chosen so that a
certain colour corresponds approximately to the density level,
corrected by the linear growth factor for that redshift. Wavelet blue
colours correspond to under-dense regions of density waves, green
colours to slightly over-density regions, and red colours to highly
over-density regions.

\begin{figure*}[ht]
\centering
\resizebox{0.9\textwidth}{!}{\includegraphics*{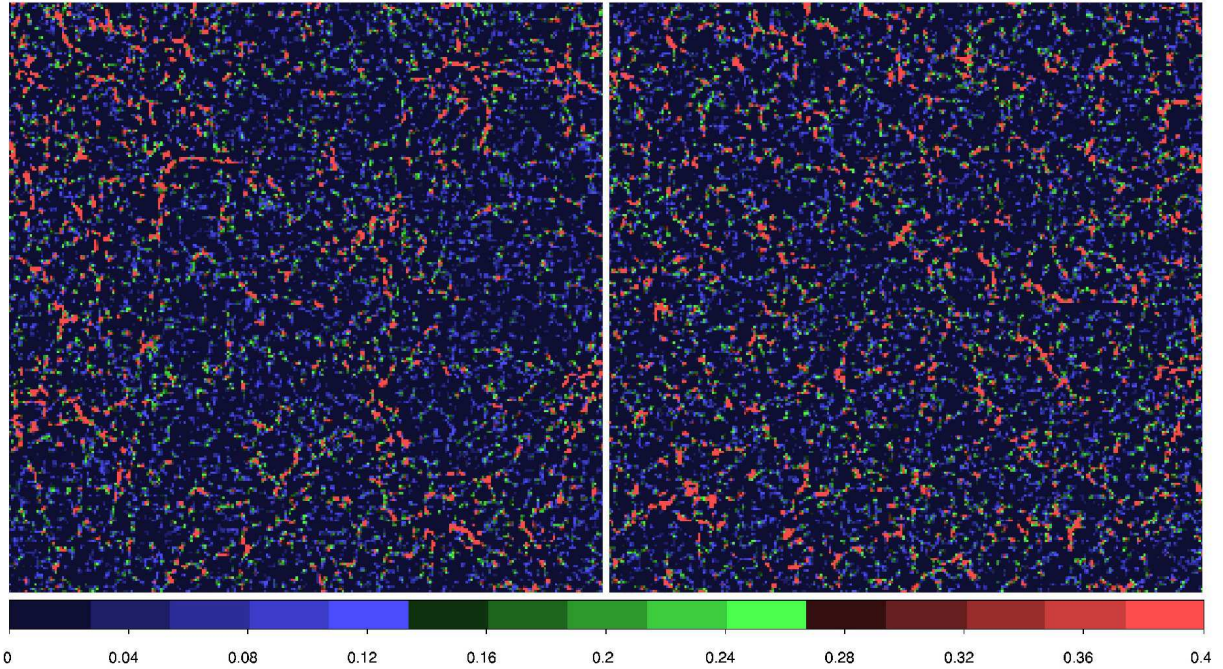}}
\caption{{Wavelets $w1$ of models L100.100 and L100.008 at
    redshift $z=10$ are shown in the left and right panels,
    respectively, at coordinate $k=51$. Densities are expressed on a
    linear scale, only over-density regions are shown.  As in
    Fig.~\ref{fig:L100den}, the central $256 \times 256$ pixels of the
    full models are shown.  The characteristic scale of density
    perturbations corresponding to this wavelet is 0.4~\Mpc.  Note the
    weakening of peak densities of the model L100.100 in the region of
    future large under-dense region, seen in Fig.~\ref{fig:L100den}.} }
\label{fig:L100w1}
\end{figure*}

Figure~\ref{fig:M256wav} shows that the pattern of the cosmic web on
wavelet $w5$ is almost identical at all redshifts, only the amplitude of
the density waves increasing approximately in proportion to the linear
growth factor. This linear growth is expected for density waves of
large scales, which are in the linear stage of growth. The pattern of
the web of the wavelet $w4$ changes little, but the growth of the
amplitude of density waves is more rapid.  The pattern of the wavelet
$w3$ changes much more during the evolution, and the amplitude of
density waves increases more rapidly, but essential features remain
unchanged, i.e.  the locations of high-density peaks and low-density
depressions are almost independent of the epoch.

Fig.~\ref{fig:M256wav} shows that at all redshifts high-density peaks
of wavelets of {\em medium and large} scales almost coincide.  In
other words, density perturbations of medium and large scales have a
tendency of phase coupling or synchronisation at peak positions.
\citet{Einasto:2011} reached the same conclusion using models with a
much broader scale interval.  Figure~\ref{fig:M256wav} shows that the
synchronisation of medium and large scales applies also to under-dense
regions.  The analysis below describes the differences between the
synchronisation of over- and under-dense regions in quantitative
terms.

\subsection{The evolution of the fine structure in the density field}

To follow the evolution of the fine structure in the density field, we
compare high-resolution density fields of the models of L100 series
(see Fig.~\ref{fig:L100den}).  We show slices at the coordinate $k=51$
of the full model L100.100, and of the strongly cut model L100.016, at
epochs $z=0$, and $z=2$.  The $k$ coordinate is chosen so that the
slice in the model L100.100 crosses a large under-dense region between
a rich supercluster and several rich clusters.  To see more clearly
the differences between the density fields of models L100.100 and
L100.016, we show in Fig.~\ref{fig:L100den} only the zoom-in of the
central $50 \times 50$~\Mpc\ ($256 \times 256$ pixels) region.  To
compare the present field with the initial density field, we use the
smallest scale wavelet $w1$ for both models at the redshift $z=10$,
shown as a zoom-in plot in Fig.~\ref{fig:L100w1}, which is similar to
the plot for the present epoch in Fig.~\ref{fig:L100den}.

The power spectrum of density perturbations has the highest power on
small scales. Thus, the influence of small-scale perturbations
relative to large-scale perturbations is strongest in the early period
of evolution of structure.  For this reason, density fields and
wavelets $w1$ at early epochs are qualitatively rather similar for the
full model L100.100, and for the model cut on small scales,
$\lambda_{\mathrm{cut}} = 8$~\Mpc, called L100.008 (see
Fig.~\ref{fig:L100w1}).

\begin{figure*}[ht]
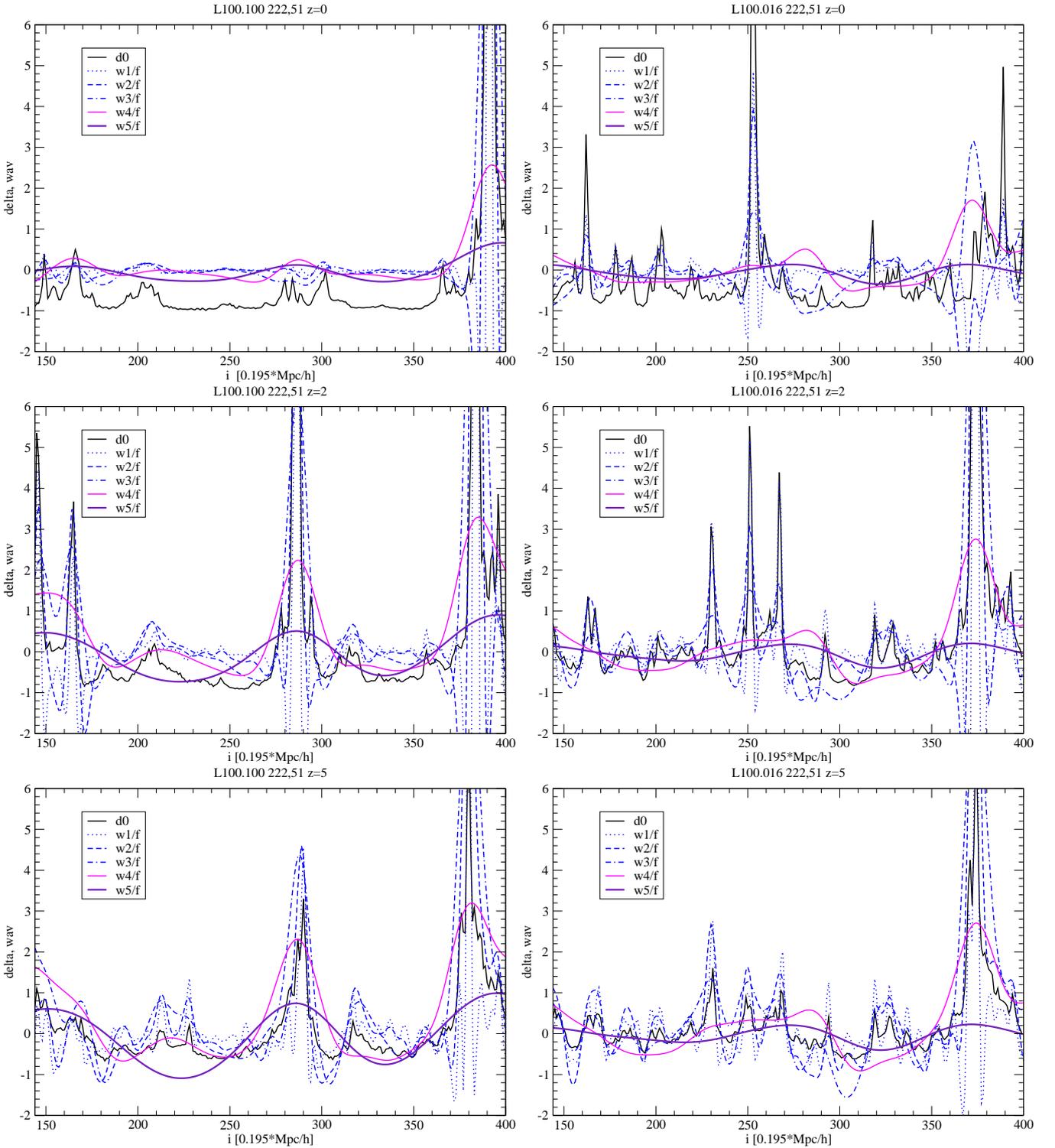

\centering
\resizebox{0.48\textwidth}{!}{\includegraphics*{17248_fig5a.eps}}
\resizebox{0.48\textwidth}{!}{\includegraphics*{17248_fig5b.eps}}\\
\resizebox{0.48\textwidth}{!}{\includegraphics*{17248_fig5c.eps}}
\resizebox{0.48\textwidth}{!}{\includegraphics*{17248_fig5d.eps}}\\
\resizebox{0.48\textwidth}{!}{\includegraphics*{17248_fig5e.eps}}
\resizebox{0.48\textwidth}{!}{\includegraphics*{17248_fig5f.eps}}\\
\caption{The evolution of the local density and wavelets of the models
  L100.100 (left panels) and L100.016 (right panels) in beams along
  the $i-$coordinate at $j=220$, $k=51$. The same $k-$coordinate was
  used in plotting the density field in Fig.~\ref{fig:L100den}.  Data
  are shown for epochs $z=0,~2,~5$. To see better details only the
  region $144 \leq i \leq 400$ of length 50~\Mpc\ is shown. The
  characteristic scale of the wavelet $w5$ is 12.5~\Mpc.  Wavelets are
divided by the  factor $f \propto (1 + z)^{-1}$.  }
\label{fig:L100beam}
\end{figure*}

However, there are small but important differences in the patterns of
small-scale structures in the models L100.100 and L100.008 at $z=10$.
The density peaks of the model L100.008 have more or less equal
heights throughout the whole simulation box, whereas in the model
L100.100 in regions of future voids the peak heights are lower than in
the future supercluster regions.  This shows that already at the epoch
$z=10$ large-scale perturbations started to influence the density
field on small scales.

\begin{figure*}[ht]
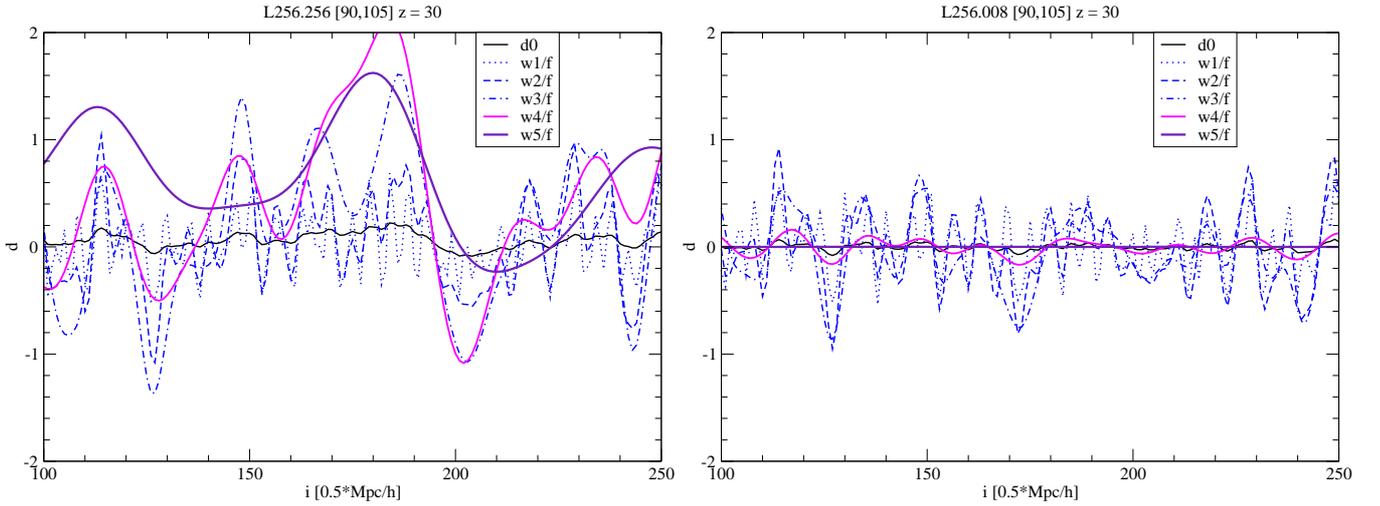

\centering
\resizebox{0.48\textwidth}{!}{\includegraphics*{17248_fig6a.eps}}
\resizebox{0.48\textwidth}{!}{\includegraphics*{17248_fig6b.eps}}\\
\caption{The local density and wavelets of the models L256.256 (left
  panels) and L256.008 (right panels) in beams along the
  $i-$coordinate at $j=90$, $k=105$ at the early epoch, $z=30$. To
  enhance details, only the region $100 \leq i \leq 250$ of length
  75~\Mpc\ is shown. The characteristic scale of the wavelet $w5$ is
  32~\Mpc.}
\label{fig:L256beam}
\end{figure*}

The colour coding is identical in all panels of
Figs.~\ref{fig:L100den}. Codes are chosen so that the density level
$D_{\mathrm{loc}} = 1.6$ is clearly visible.  Particle clouds of
density above this limit can form collapsed haloes during the Hubble
time, and below this limit they stay in pre-galactic more diffuse form.
The actual collapse parameter is 1.69 \citep{Kaiser:1984,
  Bardeen:1986}.  In both models, there exists a network of
filaments.  We see that in all models shown in Fig.~\ref{fig:L100den}
the majority of filaments have densities below this limit, i.e.  that
filaments consist of strings of primordial matter that have not yet
formed compact haloes.

\begin{figure*}[ht]
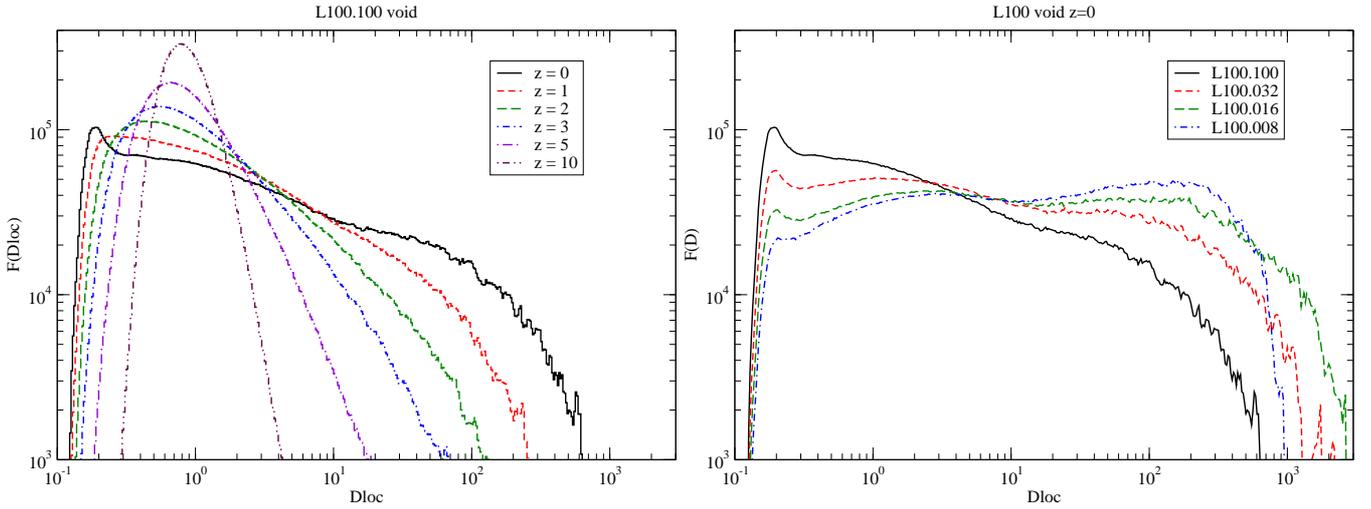

\centering
\resizebox{0.48\textwidth}{!}{\includegraphics*{17248_fig7a.eps}}
\resizebox{0.48\textwidth}{!}{\includegraphics*{17248_fig7b.eps}}
\caption{Left panel: the distribution of void particle local densities
  of the model L100.100 as a function of redshift. Right panel: the
  distributions of void particle local densities at the present epoch
  of the models of the series L100 with different cutoff
  wavelengths. }
\label{fig:voidpart}
\end{figure*}

At the epoch $z=2$, there are already strong differences between the
density fields of models L100.100 and L100.016. In the model L100.016,
the high-density knots are distributed more uniformly, and the whole
pattern of filaments has a smaller scale.  We discuss the distribution
of mean void sizes for both models below.  In the full model L100.100,
the fraction of particles above the critical density $D_{\mathrm{loc}}
= 1.6$ in regions of low global density (voids) is lower (see
Fig.~\ref{fig:voidpart}).  Regions containing high-density knots
(marked by green and red) start to form superclusters. In the model
L100.016, these regions are more uniformly distributed.

The comparison of distributions for the epochs $z=2$ and $z=0$ shows that
filaments contract with time, as known from many earlier studies cited
above.  In the model L100.016,  the pattern of
filaments changes very little between the epochs $z=2$ and $z=0$.
In contrast, for the model L100.100 density evolution can be clearly 
seen.  At $z=2$, small-scale filaments fill almost the whole space of
voids between rich systems. At $z=0$, most of these small filaments
have merged leaving more space for very low-density regions.

The differences between the models L100.100 and L100.016 at the
present epoch $z=0$ are very well seen in Fig.~\ref{fig:L100den}. In
the model L100.100, between the supercluster at the left corner and
the cluster at the right edge there is a large low-density region.
This region is crossed near the center by a filament that has several
knots in the green and red colour.  In the model L100.016, there are no
rich superclusters, the whole region being covered by a web of
small-scale filaments. In other words, large-scale perturbations 
present in the model L100.100, have suppressed the growth of the
density of filaments in void regions.

There exists a low-density smooth background, seen in the
Fig.~\ref{fig:L100den} in deep dark-blue colour.  The density of this
background, $D \approx 0.1$, is lower at the present epoch $z=0$,
i.e. the density of the smooth background decreases with time (see
Fig.~\ref{fig:voidpart}).  Regions of very low density have much larger sizes in the
model L100.100 than in the model L100.016.

\subsection{The evolution of density perturbations of various scales}

We now follow the evolution of density perturbations of various scales
in the models of the series L100 and L256 in more detail.  The
evolution of density perturbations has several characteristics: the
shapes of density perturbations of various scales and their change
with time; synchronisation, amplification, and suppression of density
perturbations of various scales; and the formation of regions of very
low density.

To see the evolution of the density field and its wavelets, we show in
Fig.~\ref{fig:L100beam} one-dimensional density distributions (beams)
along horizontal lines of the plane in Fig.~\ref{fig:L100den}. The
distributions are generated along the axis $j=222$, $k= 51$, for the
models L100.100 and L100.016 at the redshifts 0,~2,~5.  Beams are
taken along the $i-$coordinate; at each $i$ value, all cells in the
$j-$ and $k-$coordinate within $\pm 5$ from the centre of the beam are
counted, i.e. beams have sizes $11\times 11$ cells (about $2.15 \times
2.15$~\Mpc\ in the models of the L100 series). The average densities
for a given $i$ are found, and are given in the mean density units.
We use overdensities $D-1$ here, so the mean overdensity level is
zero, which is similar to the mean value for wavelets.  Wavelet
amplitudes are divided by the linear growth factor, thus during linear
growth their amplitudes should not change.  The effective scale for
the wavelet $w5$ of the model L100 is 12.5~\Mpc, as seen also from the
separation between the high-density peaks in Figs.~\ref{fig:L100beam}.

At high redshifts, perturbations of various scales have almost
sinusoidal shapes, and the wave peaks have approximately equal
heights.  During the subsequent stages of evolution,
\citet{Einasto:2011} showed that perturbations of larger scales begin
to affect the evolution. These perturbations amplify small-scale
perturbations near maxima, and suppress small-scale perturbations near
minima.  Thereafter, still larger perturbations amplify smaller
perturbations near their maxima, and suppress smaller perturbations
near their minima, and so on.

For early stages, the density fields and wavelets are given for the
models L256.256 and L256.008, for the epoch $z=30$, in
Fig.~\ref{fig:L256beam}.  This model also allows us to illustrate the
influence of larger waves, because the characteristic scale for the
wavelet $w5$ of the model L256 is 32~\Mpc.

We see that in the model L256.008 wavelets of different order have
approximately sinusoidal shapes.  The wavelet $w5$ of this model has
zero amplitude, and the amplitude of the wavelet $w4$ is small,
because in this model only waves of scale up to 8~\Mpc\ are present.
The wavelet $w3$ of the effective scale, which is approximately equal
to the cut scale of this model, has the highest amplitude.  The maxima
and minima of this wavelet are partly synchronised with the maxima and
minima of the wavelets $w2$ and $w1$; synchronisation is better in the
regions that happen to coincide with the maxima and minima of the
wavelet $w4$.

In the model L256.256 at the early epoch $z=30$, the wavelet $w5$ has
the highest amplitude, the next highest being that of the wavelet
$w4$.  These two wavelets are only partly synchronised, i.e. the
maxima of the wavelet $w5$ coincide in position only approximately
with the maxima of the wavelet $w4$, and near the minima of the
wavelet $w5$ there is a secondary maximum of the wavelet $w4$,
i.e. the wavelet $w4$ behaves as the first overtone of the wavelet
$w5$.  The shape of small-scale wavelets is not sinusoidal.  This is
caused by large-scale waves that have started to change the shapes of
small-scale waves.

Further evolution with time of the wavelets can be followed using the
model L100 in Fig.~\ref{fig:L100beam}.  We discuss first the evolution
of wavelets in the model L100.016 (see the right panels of the
Figure).  The largest wavelet $w5$ has for all redshifts approximately
the same amplitude and a sinusoidal shape, suggesting that density
perturbations of this scale are in the linear growth regime.  The
shape of the next wavelet $w4$ is very different from sinusoidal.  In
some regions of maxima of the wavelet $w5$, the wavelet $w4$ has very
strong maxima. An example is the region near $i \approx 370$, where
all wavelets of smaller scale also have strong maxima, and wavelets of
all scales up to $w4$ are very well-synchronised.  The overall shape
of the density profile is determined by the wavelet $w3$ that has a
characteristic scale 3.1~\Mpc.  Most peaks seen in the density profile
are due to the maxima of this wavelet.  In most of these peaks,
wavelets of smaller scale also have maxima, i.e. near the peaks
small-scale wavelets are synchronised.

When we compare the density and wavelet distributions of this model at
various epochs, we see little difference.  This shows that in the
model L100.016 the structure has rapidly evolved at early epochs and
changes only a little later.  The most important development is the
decrease in the density in deep void regions, and the increase in the
density in massive haloes.

We now consider the evolution of the full model L100.100, shown in the
left columns of Fig. ~\ref{fig:L100beam}.  We see that the evolution
of the largest wavelet $w5$ is almost linear up to the epoch $z \geq
1$ (the density and wavelet distributions for the epochs $z=1$ and
$z=2$ are rather similar, only the amplitudes of wavelets up to $w4$
are smaller for $z=1$). The shapes of the wavelet $w5$ for different
epochs are almost sinusoidal and the heights of the maxima are
approximately equal.  The next wavelet $w4$ behaves as a first
overtone of the wavelet $w5$ -- near the minima of $w5$ there are
maxima of $w4$, which have much lower heights than the maxima near the
maxima of $w5$. Evidence of this phenomenon can also be clearly seen
in the wavelet analysis of the Sloan Digital Sky Survey (see Fig.~6 of
\citet{Einasto:2011}).  Near the joint maxima of $w5$ and $w4$, there
are very strong maxima of all wavelets of smaller order; this is very
well seen at locations $i \approx 290$ and $i \approx 380$.

Near the minima of $w5$ and the maxima of $w4$, wavelets of smaller
order also have maxima, but these maxima get weaker at lower
redshifts, as in the region around $i \approx 220$. Here, small-scale
wavelets are partly synchronised, and small-scale peaks of the density
field are related to maxima of the wavelets $w2$ and $w3$.

The density and wavelet distributions of the model L100.100 for the
present epoch $z=0$ are completely different from the distributions at
higher redshifts.  In the present epoch the dominant feature is a
large under-dense region in the interval $120 < i < 380$.  This large
under-dense region is caused by large-scale density perturbations that
are not shown as wavelets in the Figure.  In this region, these
large-scale density waves have their minima and the amplitudes of
density waves of smaller scales, including $w4$ and $w5$, are 
suppressed.  The amplitudes of wavelets $w3$ and lower orders are
almost zero.  Near the density maxima seen at higher redshifts at $i
\approx 210$ and $i \approx 290$, there are very weak density peaks
with maxima below the mean density level.  These maxima are seen in
the density field as weak filaments (see Fig.~\ref{fig:L100den}).

The most remarkable feature of the density field in the model L100.100
at the present epoch is the presence of large under-dense regions of
very low density $D \approx 0.1$, which can be clearly seen in
Fig. ~\ref{fig:L100den} in deep-blue colour.  At earlier epochs, the
density in these regions was higher and there were numerous
low-density peaks within the regions, for the present epoch most of
these peaks are gone.  This is caused by density perturbations on
larger scales.

When we compare the evolution of density distributions of models
L100.100 and L100.016, we see remarkable differences.  These
differences are solely due to the presence of density perturbations of
scales larger than $\lambda_{\mathrm{cut}} = 16$~\Mpc\ in the model
L100.100.  We note that both models were generated with identical
``random amplitudes'', i.e. the perturbations of scales $\lambda \leq
\lambda_{\mathrm{cut}} = 16$~\Mpc\ are identical in both models.

The main result of the paper by \citet{Einasto:2011} was that at all
redshifts high-density peaks of wavelets of large and medium scales
almost coincide.  Figures~\ref{fig:M256wav}, \ref{fig:L100den}, and
\ref{fig:L100beam} show that the same conclusion is valid for
positions of density depressions (deepest voids) of wavelets of large
and medium scales.  The other main conclusion of \citet{Einasto:2011}
was that positions of peaks of waves of different scale coincide,
i.e. density waves of different scale are synchronised.  A look at
Figs.~\ref{fig:M256wav} and \ref{fig:L100beam} shows that the
synchronisation of density waves of different scales also concerns
density depressions of large and medium scales.  However, the
synchronisation of the depressions of density waves, which is
responsible for the formation of voids, is less pronounced than that
of density peaks.

\subsection{The evolution of density distributions in void  regions}

To investigate the evolution of the density distribution in void
regions, we selected in all models void particles in the present epoch
$z=0$. We calculated the distributions of the global densities of full
models (i.e. models with no cuts in the power spectra), and selected
particles of the lowest global density values, about 10~\%\ of all
particles  (13.390.895 particles in the model L100.100). The
corresponding value of the threshold global density is 0.565 for the
model L100.100 (in units of the mean density). 

The evolution of the distributions of local densities of void
particles for the full model L100.100 is given in the left panel of
Fig.~\ref{fig:voidpart}.  The distributions of void particle local
densities for the models of the L100 series for the present epoch are
shown in the right panel of Fig.~\ref{fig:voidpart}.

Figure~\ref{fig:voidpart} shows that the initial distribution of
particles at the redshift $z=10$ is quite symmetrical in the log-log
representation.  The density distribution has a peak at $D \approx
0.8$.  As time goes on, the peak density decreases, and at the present
epoch has a value $D \approx 0.2$; the lowest density occurs close to
$D \approx 0.1$.  Particles in overdense regions ($D \geq 1$) form
haloes; with decreasing redshift $z$, the peak densities increase,
i.e. haloes in void regions become more massive and denser.

Density evolution depends strongly on the cutoff wavelength of the
model.  In the most strongly cut model L100.008 at the present epoch,
the fraction of particles in very low-density regions ($D \approx
0.2$) is about five times lower than in the full model L100.100.  In
the model L100.008, most particles in void regions form haloes of mean
density $D \approx 100$, and with the maximum density $D \approx
1000$. With the increase in the cut wavelength
$\lambda_{\mathrm{cut}}$, the fraction of particles in very
low-density regions increases.  The maximal density of haloes in void
regions reaches the highest value, $D \approx 2500$, in the model
L100.016.  In this model, density waves between the scales 8 and
16~\Mpc\ amplify the density of haloes.  If density perturbations of
larger scale are included (models L100.032 and L100.100), then in the
void regions still larger perturbations start to decrease the maximum
masses of haloes. This depression is largest in the full model
L100.100, where the maximum local densities in void haloes reach the
values $D \approx 600$.

\section{Correlation analysis of wavelet-decomposed density fields}

We now attempt to quantify some of the qualitative statements given
earlier in the text. To this end, we perform the correlation analysis
of wavelet-decomposed density fields. Our approach is analogous to the
one presented in \citet{Einasto:2011} with the exception that here
instead of over-densities we focus on under-dense regions. In the
following, we present our results only for the model M256 since the
other ones lead to very similar results. We consider six wavelet
levels: $w1$, $w2$, $\ldots$, $w6$ with the effective smoothing scales
of $4$, $8$, $\ldots$, $128$~\Mpc, respectively, and use the simulated
density fields at five different redshifts $z=30$, $10$, $5$, $1$,
$0$.

Quite generally, one can choose two redshifts, $z_i$ and $z_j$, along
with two wavelet levels, $w_m$ and $w_n$, and calculate the
correlators
\begin{equation}
r_{w_m z_i,\,w_n z_j}=\frac{\langle (\delta_{w_m z_i}-\langle \delta_{w_m z_i} \rangle)(\delta_{w_n z_j}-\langle \delta_{w_n z_j} \rangle)\rangle}{\sqrt{\langle(\delta_{w_m z_i}-\langle \delta_{w_m z_i} \rangle)^2\rangle\langle(\delta_{w_n z_j}-\langle \delta_{w_n z_j} \rangle)^2\rangle}},
\label{eq:correlators}
\end{equation}
where $\delta_{w_m z_i}$ corresponds to the wavelet-decomposed density
field for level $w_m$ at redshift $z_i$. The angle brackets represent
an ensemble average, which under the ergodicity assumption is replaced
by a simple spatial average. We note that we calculate the correlators
for zero lag only, i.e. we do not shift one field with respect to the
other.

\begin{figure}[ht]
\centering
\includegraphics[width=0.5\textwidth]{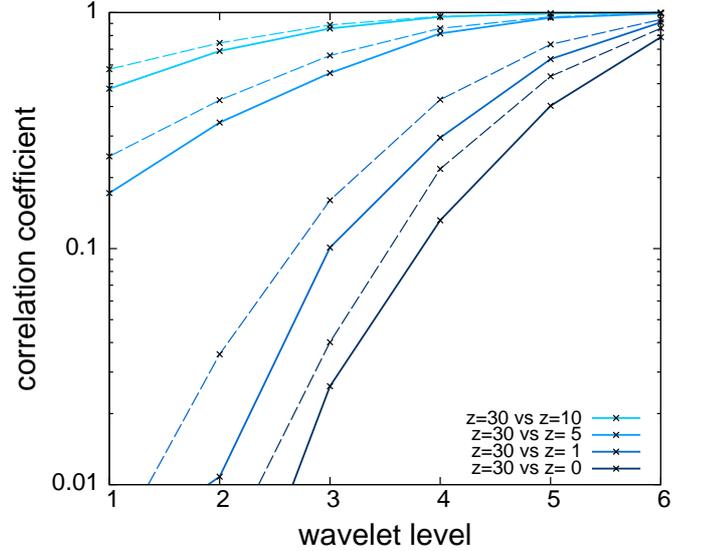}
\caption{The behaviour of the correlation coefficient for different
  redshift pairs $(z_i,z_j)\in \{(30,10);(30,5);(30,1);(30,0)\}$ for
  all the six wavelet levels. The solid and dashed lines correspond to
  the under- and overdensities, respectively ($10\%$ of the most
  under/over-dense cells). We see that for the largest smoothing scale,
  i.e. $w6$, all the correlators stay quite close to $r=1$, while
  later on, as the other redshift decreases below $z=30$, the lines
  start to deviate from $r=1$. Thus, on the largest scales the
  information is approximately preserved, while on the smallest scales
  the information gets gradually erased.}
\label{fig:correlate:fixed_w}
\end{figure}

\begin{figure}[ht]
\centering
\includegraphics[width=0.5\textwidth]{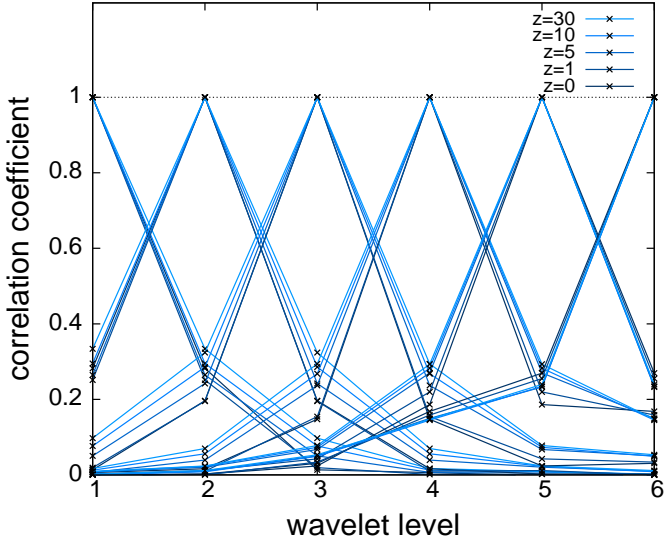}
\caption{Correlators at fixed redshifts $z_i=z_j=30,10,5,1,0$ from top
  to bottom (from light to dark blue) for the under-dense regions. For
  $w_m=1$, the curves are peaked at $w_n=1$, while they drop gradually
  as $w_n$ is increased to higher values. Similarly, the curves for
  $w_m=2$ are peaked at $w_n=2$, and get reduced as the distance
  increases from this point. For the other $w_m$ and $w_n$ values, the
  behaviour is very similar. We see that the lower the output
  redshift, the narrower  the coupling kernels, i.e. a nonlinear
  evolution in under-dense regions leads to additional decoupling of
  nearby wavelet modes.}
\label{fig:correlate:fixed_z}
\end{figure}

In what follows, the above correlators are not calculated for the full
density fields, but instead masks are applied  to separate 
under-dense void regions. If two wavelet fields $\delta_{w_m z_i}$ and
$\delta_{w_n z_j}$ are correlated then the mask is always defined with
the field that has larger smoothing scale, e.g., the field $\delta_{w_m
  z_i}$ if $w_m\ge w_n$. The masking level is taken such that only
$10\%$ of the most under-dense cells end up inside the mask.

In the following, we  use two types of correlators:
\begin{enumerate}
\item Fixed wavelet scale correlators, i.e. $w_m=w_n$, at different redshifts.
\item Correlators at fixed redshifts, i.e. $z_i=z_j$, for different wavelet levels.
\end{enumerate}

In the first case, we take one of the density fields always at
redshift $z=30$, which is high enough for all of the scales of
interest to be well in the linear regime. It is easy to understand
that under the linear evolution, where the values of $\delta$ just get
multiplied by the same scale-independent but time-varying factor, the
correlation coefficient should always stay at the value $r=1$,
i.e. all of the initial information is well preserved. In
Fig.~\ref{fig:correlate:fixed_w}, we show the behaviour of the
correlation coefficient for different redshift pairs: $(z_i,z_j)\in
\{(30,10);(30,5);(30,1);(30,0)\}$ for all the six wavelet levels. It
is easy to see that for the largest smoothing scale, i.e. $w6$, all
the correlators stay rather close to $r=1$, while later on, as the
other redshift gets smaller than $z=30$ the lines start to decline
from $r=1$, especially on the smallest scales. Thus, on the largest
scales the information is approximately conserved, while on the
smallest scales the information gets erased. The lower the redshift of
the other density field the greater the loss of information.  In
practice, for the cases $z=10$ and $z=5$ the loss of information is
relatively modest if the wavelet level $w\ge 3$. For $z=1$ and $z=0$,
the information is approximately saved only for the largest
scales. The dashed lines in Fig.~\ref{fig:correlate:fixed_w} show the
corresponding results for the over-dense regions, in this case
focusing on the $10\%$ of the most over-dense cells. As one can see,
the information loss in under-dense regions occurs more rapidly with
time than for the over-densities.

In Fig.~\ref{fig:correlate:fixed_z}, we plot the correlators at fixed
redshifts $z_i=z_j=30,10,5,1,0$ from top to bottom (from light to dark
blue). For $m=1$, the curves are peaked at $n=1$, while their
amplitude decrease gradually as $n$ increases to higher
values. Similarly, the curves for $m=2$ are peaked at $n=2$, and
decrease in amplitude as one moves to neighbouring wavelet levels. For
the other $n$ values, the behaviour is very similar. As long as the
evolution proceeds in a linear manner, i.e. the growth depends only on
redshift, but is independent of the wavelet scale, the coupling
kernels plotted in Fig.~\ref{fig:correlate:fixed_z} should stay
exactly the same. However, we see that the lower the output redshift,
the narrower the coupling kernels, i.e. a nonlinear evolution in
under-dense regions leads to the additional decoupling of the nearby
wavelet modes.

The corresponding figure for the over-densities was given in
\citet{Einasto:2011} (see Fig. 7 there). The main difference between
Fig.~7 by \citet{Einasto:2011} and Fig. ~\ref{fig:correlate:fixed_z}
is that in the case of over-densities the coupling kernels become
broader because of nonlinear evolution, i.e. instead of
desynchronisation we have increasing synchronisation of over-densities
of different wavelet levels.

\begin{figure}[ht]
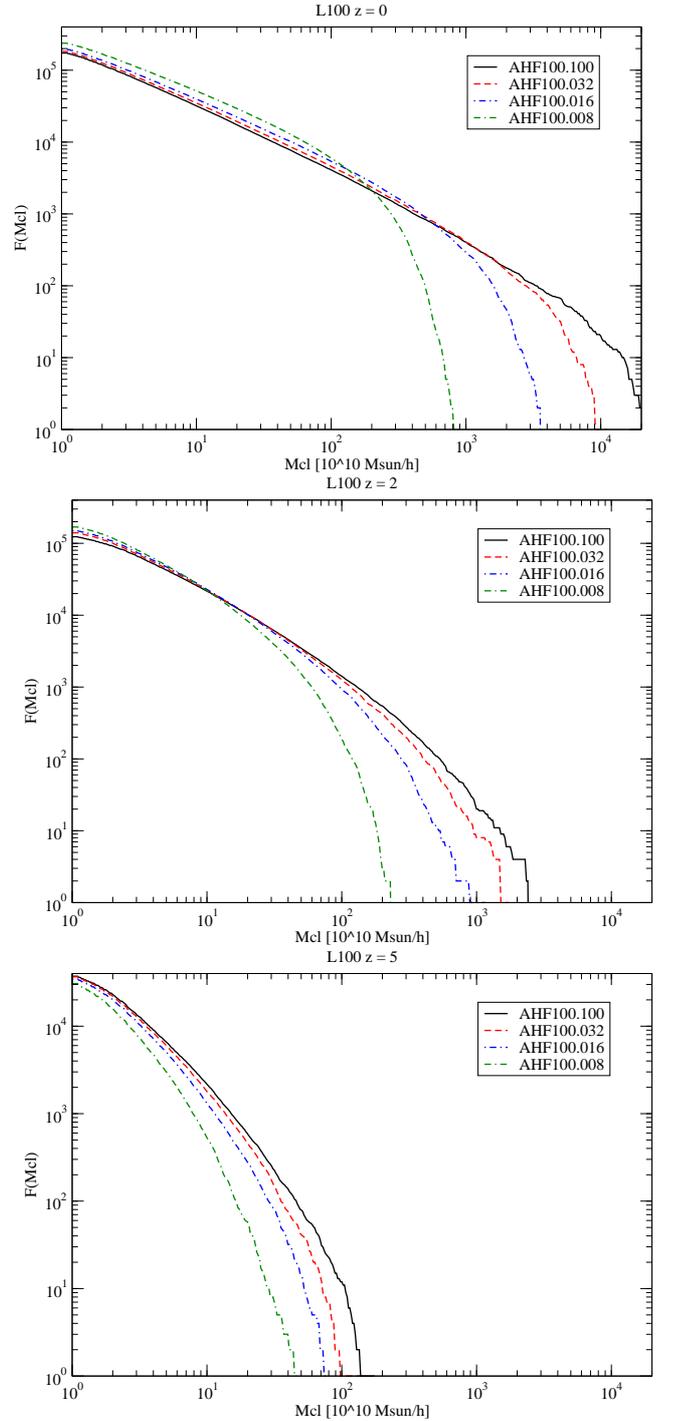
 
\centering 
\resizebox{0.45\textwidth}{!}{\includegraphics*{17248_fig10a.eps}}\\
\hspace{2mm}  
\resizebox{0.45\textwidth}{!}{\includegraphics*{17248_fig10b.eps}}\\
\hspace{2mm}  
\resizebox{0.45\textwidth}{!}{\includegraphics*{17248_fig10c.eps}}\\
\hspace{2mm}  
\caption{The cumulative mass functions of the AHF haloes of models of
  the L100 series with various cutoff scales.  The upper, middle, and
  lower panels are for the redshifts $z=0,~2,~5$, respectively.  }
\label{fig:L100_AHFmassf} 
\end{figure} 

It is important to realise that even with only the linear evolution of
the Gaussian density field the nearby wavelet levels at fixed redshift
get significantly coupled, since the neighbouring levels tend to
contain some of the common Fourier space modes. However, assuming only
linear evolution it is clear that the coupling does not change with
redshift.

\begin{figure*}[ht]
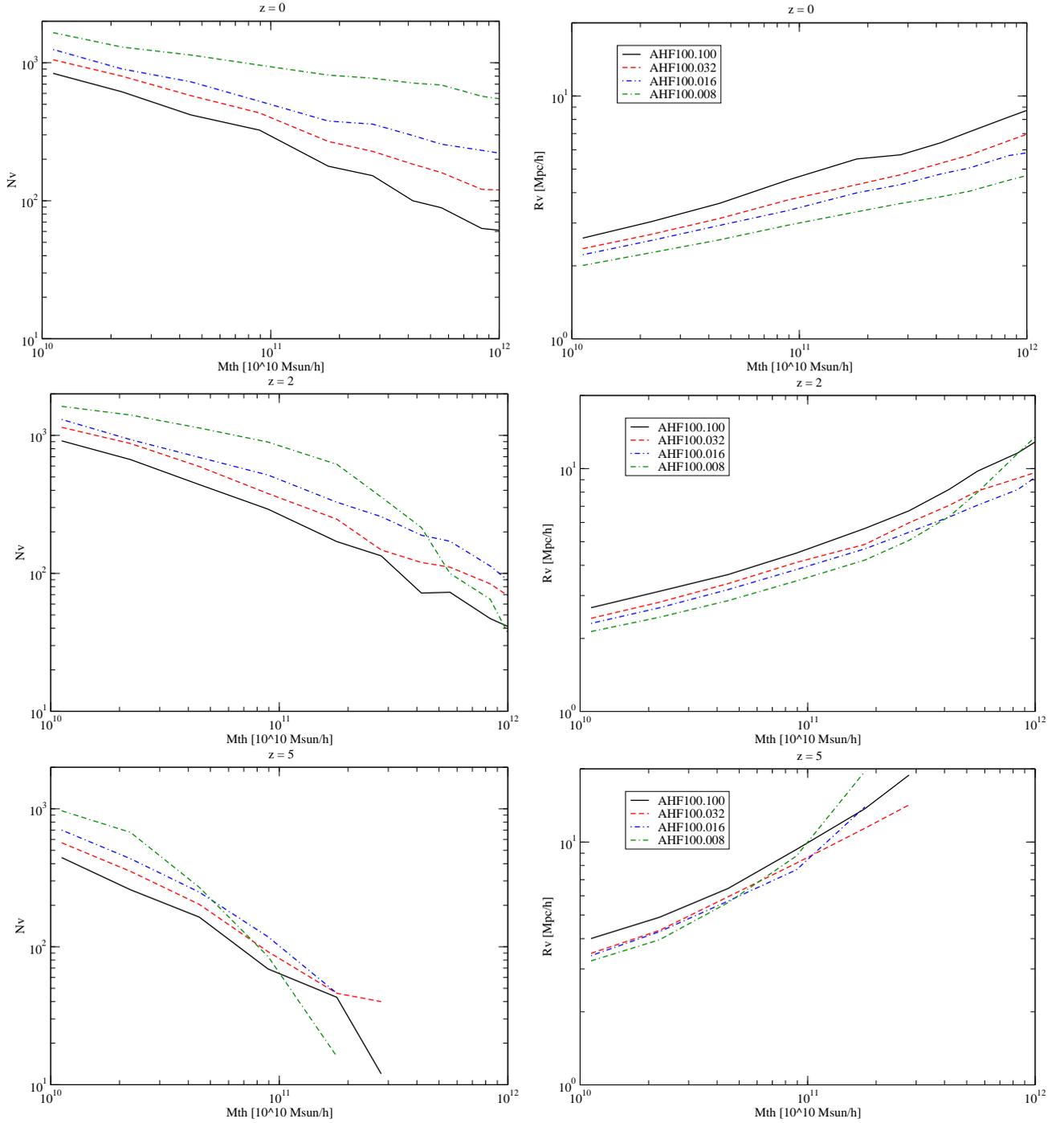
 
\centering 
\resizebox{0.45\textwidth}{!}{\includegraphics*{17248_fig11a.eps}}
\hspace{2mm} 
\resizebox{0.45\textwidth}{!}{\includegraphics*{17248_fig11b.eps}}\\
\hspace{2mm} 
\resizebox{0.45\textwidth}{!}{\includegraphics*{17248_fig11c.eps}}
\hspace{2mm} 
\resizebox{0.45\textwidth}{!}{\includegraphics*{17248_fig11d.eps}}\\
\hspace{2mm} 
\resizebox{0.45\textwidth}{!}{\includegraphics*{17248_fig11e.eps}}
\hspace{2mm} 
\resizebox{0.45\textwidth}{!}{\includegraphics*{17248_fig11f.eps}}\\
\hspace{2mm} 
\caption{The left panels show the numbers of voids, defined by the AHF haloes for
  various threshold masses and models, as shown in Fig.~\ref{fig:L100_AHFmassf}.  The
  right panels show the mean radii of voids, defined by the AHF haloes
  for different threshold masses, $M_{\mathrm{th}}$, and for various
  cut-off scales. The upper, middle, and lower panels are for the redshifts
  $z=0,~2,~5$, respectively.  }
\label{fig:L100_AHFvoids} 
\end{figure*} 

\section{The structure of voids}

Cosmic voids are defined by objects surrounding them -- galaxies and
clusters of galaxies of various luminosity (mass).  In the case of
models, it is customary to use dark matter haloes instead of galaxies
or clusters.  To investigate the influence of density perturbations of
various scale on the void structure, we shall use our models of the
L100 series which have the highest resolution in mass and scale.  To
find haloes, we applied the Amiga halo finder by
\citet{Knollmann:2009}.  We characterise the web and void structure by
the dependence of the halo mass function on the scale of density
perturbations for various epochs, and by the number and radii of voids
defined by haloes of various mass. Our results are shown in
Figs.~\ref{fig:L100_AHFmassf} and \ref{fig:L100_AHFvoids}.

Fig.~\ref{fig:L100_AHFmassf} shows the cumulative mass functions of
the AHF haloes for all models of the L100 series, for three epochs,
$z=0,~2,~5$.  We see that in the models where the large-scale waves
have been cut off, maximum haloes have much lower masses than in the
full models.  This effect can also be seen in Fig.~\ref{fig:L100den}.
The differences between the models with various cutoff scales increase
with time: at early epochs, halo masses are lower and time is needed
for most massive haloes to grow.

To find voids, we used a simple void finder proposed by
\citet{Einasto:1989a}.  For each vertex of the simulation grid, we
first calculated its distance to the nearest AHF halo.  The maxima of
the void distance matrix correspond to the centres of voids, and their
values are the void radii.  The distribution of the AHF haloes is
noisy, thus there are many nearby local maxima in the distance matrix.
We define the position of the void centre as the location of the cell,
which has the largest distance to a halo within a box of the size of
$\pm 3$ grid elements.

The left panels in Fig.~\ref{fig:L100_AHFvoids} show the number of
voids found for various AHF halo mass thresholds, which correspond to
systems of galaxies of different mass.  The right panels of the Figure
show mean radii of voids as a function of the AHF halo mass threshold.
The void numbers and radii characterise the hierarchy of voids.  We
see that as the threshold halo mass used in the void search increases,
the number of voids continuously decreases, and the void radii
increase.  This means that some filaments are fainter than the
respective mass threshold limit, and do not contribute to the void
definition.  Both parameters also depend strongly on the highest
density perturbations used in the models.  Models cut at larger scales
have more voids, but their radii are smaller.  This effect can be
clearly seen at all simulation epochs.

In the models L100.100 and L100.032, the dependence of the number of
voids and their mean radii is a more-or-less continuous function of
the halo threshold mass. In contrast, in the strongly cut models
L100.016 and L100.008, at higher mass thresholds, the number of voids
decreases very rapidly and void radii also increase more rapidly than
in the models L100.100 and L100.032.  This effect is due to the very
sharp decrease in the number of haloes of high mass (see
Fig.~\ref{fig:L100_AHFmassf}).  These rare haloes define very large
voids.  The sizes of these voids are not characteristic of the overall
cosmic web pattern of the particular model.
 
The simulation box used in the present void structure study has the
size $L=100$~\Mpc, thus very large perturbations responsible for the
formation of rich superclusters are absent in the model.  For this
reason, the largest voids for the highest halo mass thresholds have
radii $\simeq 10$~\Mpc.  As shown by \citet{Suhhonenko:2011}, in
models with cube sizes $L=265$ and $L=768$~\Mpc\ the maximum void
radii are much larger (see Fig.~6 by \citet{Suhhonenko:2011}).  This
comparison shows that large-scale density perturbations are needed to
form voids defined by superclusters.

\section{Discussion}

According to the current cosmological paradigm, the cosmic web with
systems of galaxies of various scale and mass, from clusters to
filaments and superclusters, and voids between them, is formed from
tiny density perturbations during the very early stage of the
evolution of the Universe.  For the formation of the web and of voids
between various objects of the web, the presence of a continuous
spectrum of density perturbations of various scales is essential.  The
power spectrum of density perturbations has the highest power on small
scales. Thus, the influence of small-scale perturbations relative to
large-scale perturbations is strongest in the early period of
structure evolution.  Small-scale systems, i.e. small haloes in
simulations and dwarf galaxies in the real world are the earliest
compact objects to form. As shown by \citet{Suhhonenko:2011}, and
confirmed in the present study, small-scale haloes form at early
epochs everywhere. The wavelet analysis done by \citet{Einasto:2011}
shows that wavelets $w1$ at redshift $z=30$ are almost identical in
models L256.256 and L256.008 (see Fig. 8~there). The present study
shows that this is also the case for the wavelets $w1$ of the models
L100.100 and L100.008 at the epoch $z=10$ (see Fig.~\ref{fig:L100w1}).

Further evolution of the web depends on the presence of density
perturbations on larger scales.  In models where medium-scale density
perturbations are absent, no systems of filaments and voids form
\citep{Suhhonenko:2011}. In models with density perturbation spectra
cut on large scales, the cosmic web with filaments and voids has a
characteristic scale of the largest scale present in the density
perturbation field.  The main quantitative characteristics of the web
-- the masses of haloes, the number and sizes of voids defined by
haloes of various mass -- depend on the largest scale perturbations
present.

In the models with strongly cut power spectra ($\lambda_{\mathrm{cut}}
\leq 16$~\Mpc), the maximum masses of haloes are lower than in the
models with a full power spectrum. Their number is larger, but they
define a cellular cosmic web with smaller mean void sizes (see
Fig.~\ref{fig:L100_AHFvoids}).  With the increase of the cut-off
wavelength $\lambda_{\mathrm{cut}}$, the maximum masses of haloes
increase, and they define a cellular web with larger cells but fewer
voids.

The wavelet analysis described in previous Sections shows a very
important property of the evolution of density waves with time: the
synchronisation of the phases of density waves on various scales.
\citet{Einasto:2011} discussed this property of the evolution of
over-density features -- clusters and superclusters of galaxies in
numerical simulations.  In the present paper, we have followed the
evolution of both over- and under-density regions. The analogy in the
evolution of over- and under-density regions is expected, since in the
early linear stage of the evolution of structure positive and negative
parts of density waves were similar and symmetrical.

The wavelet analysis leads us to the conclusion that the properties of
the large-scale cosmic web with filaments and voids depend on two
connected properties of the evolution of density perturbations. The
first property is the synchronisation of density waves of medium and
large scales.  Due to the synchronisation of density waves of
different scales, positive amplitude regions of density waves add
together to form rich systems of galaxies, and negative amplitude
regions of density waves add together to decrease the mean overall
density in voids.  The amplification of density perturbations is
another property of density evolution.  Due to the addition of
negative amplitudes of medium and large scale perturbations, there is
no possibility for the growth of the initial small-scale positive
density peaks in void regions. For this reason, small-scale
protohaloes dissolve there. In the absence of medium and large-scale
density perturbations, these peaks would contract to form haloes,
which would also fill the void regions, i.e. there would be no void
phenomenon as observed.

Simulations with truncated power spectra were performed by
\citet{Little:1991} and \citet{Einasto:1993}.  \citet{Little:1991}
used three-dimensional simulations of resolution $128^3$ with power
spectra in the form $P(k) \sim k^{-1}$ for $k\leq k_c$ and $P(k) = 0$
for $k > k_c$, i.e. the spectra were cut at {\em small} scales, in
contrast to our study here. The cuts were scaled so that $k=1$
represents the fundamental mode of the simulation box of the size
$L=64$~\Mpc. The authors used the cuts: $k = 2,~4,~8,~16,~32,~64$.
The main result of the study was that the structure of the cosmic web
depends on density perturbations {\em of larger scale} than the
cut-off scale, in accordance with our results.  \citet{Einasto:1993}
made a two-dimensional simulation of the resolution $512^2$ with the
full power spectrum, and a power spectrum cut at the scale $\lambda_t
= L/4$, where $L$ is the size of the simulation box.  The fine
structure of filaments in both models was rather similar, only the
location and strength of filaments was slightly different.  This
result is also in agreement with our present findings.

As the analysis shows, the phase synchronisation of both positive and
negative sections of density waves is stronger for density waves of
larger scales, $\lambda \geq 32$~\Mpc. Scales larger than the sound
horizon at recombination, $\approx 146$ Mpc according to the most
recent cosmological data by \citet{Jarosik:2010}, were outside the
horizon most of the time.  This scale, 105~\Mpc\ for the presently
accepted Hubble constant $h=0.72$, is surprisingly close to the
characteristic scale of the supercluster-void network
\citep{Einasto:1997a,Einasto:2001ff}.  The skeleton of the
supercluster-void network was created during the very early
post-inflation stage of the evolution of the Universe
\citep{Kofman:1988}.  This result is also true for large voids between
superclusters of galaxies -- the seeds for these supervoids were
created in the very early Universe.

\section{Conclusions}

Our present study of the evolution of density perturbations of various scales
has led to the following conclusions:

\begin{itemize}

\item{} The formation of the cosmic web with filaments and voids is
  due to the synchronisation of density waves of medium and large
  scales, and the amplification of both over- and under-dense regions.

\item{} Voids are regions in space where medium- and large-scale
  density waves combine {\em in similar under-density phases}.

\item{} Owing to  phase synchronisation, the mean density of matter in
  void regions is below the mean density, thus initial small-scale
  perturbations cannot grow.

\end{itemize}

\begin{acknowledgements}

  We thank the referee for constructive suggestions. Our special
  thanks go to Rien van de Weygaert and other participants of the
  workshop ``Cosmic Web Morphology and Topology'', held in Warsaw 12
  -- 17 July 2011, for a detailed discussion of void structure
  problems.  The present study was supported by the Estonian Science
  Foundation grants No.  7146 and 8005, and by the Estonian Ministry
  for Education and Science grant SF0060067s08. It has also been
  supported by ICRAnet through a professorship for Jaan Einasto, and
  by the University of Valencia (Vicerrectorado de Investigaci\'on)
  through a visiting professorship for Enn Saar and by the Spanish MEC
  projects ``ALHAMBRA'' (AYA2006-14056) and ``PAU'' (CSD2007-00060),
  including FEDER contributions. J.E., I.S., and E.T.  thank
  Leibniz-Institut f\"ur Astrophysik Potsdam (using DFG-grant Mu
  1020/15-1), where part of this study was performed.  J.E. also
  thanks the Aspen Center for Physics and the Johns Hopkins University
  for hospitality where this project was started and continued.  In
  plotting of density fields and wavelets, we used the SAOImage DS9
  program. A.A.S. acknowledges the RESCEU hospitality as a visiting
  professor. He was also partially supported by the Russian Foundation
  for Basic Research grant No. 11-02-00643 and by the Scientific
  Programme ``Astronomy'' of the Russian Academy of Sciences.

\end{acknowledgements}

\begin{appendix}

\section{Density field and wavelets}
\label{sec:DF}

\subsection{Density field}

For each particle, we calculated the global density at the location
of the particle.  For this purpose, we first found the high-resolution density
field, using a $B_3$ spline
\begin{equation} 
B_3(x)=\frac1{12}\left[|x-2|^3-4|x-1|^3+6|x|^3-4|x+1|^3+|x+2|^3\right];
\end{equation} 
this function  differs from zero only in the interval $x\in[-2,2]$.
The one-dimensional $B_3$ box spline kernel 
of width $h=N$ is
\begin{equation} 
K_B^{(1)}(x;N)=B_3(x/N)/N.
\end{equation} 
This kernel preserves the interpolation property (mass conservation)
for all kernel widths that are integer multiples of the grid step, $h=N$.
The 3-D $K_{B}^{(3)}$ box spline kernel we use is given by the direct 
product of the three one-dimensional kernels
\begin{equation} 
K_B(\mathbf{x};N)\equiv K_B^{(3)}(\mathbf{x};N)=K_B^{(1)})(x;N)K_B^{(1)}(y;N)
        K_B^{(1)}(z;N),
\end{equation} 
where $\mathbf{x}\equiv\{x,y,z\}$. To calculate the high-resolution density
field, we use the  kernel of scale, equal to the cell size of the
particular simulation.

\subsection{Wavelets}

We use the \'a trous wavelet transform \citep[for details
see][]{Starck:1998sy,Starck:2002lt}. The field is decomposed into
several frequency bands as follows. The high-resolution (zero level)
density field was calculated with the $B_3$ spline kernel with width
equal to the size of one cell of the field, every subsequent field
being 
calculated with a  kernel twice as wide.  Wavelets were found by
subtracting higher level density fields from the previous level
fields. In such a way, each wavelet band contains waves twice the scale
of the previous band, in the range $\pm\sqrt{2}$ centered on the mean
(central) scale.  The sum of these bands restores the original density
field.

The '\'a trous algorithm' wavelet transform decomposes an $n\times n
\times n$ data set $D$ as a superposition of the form
\begin{equation}
D = D_J + \sum_{j=1}^J w_j,
\end{equation}
where $D_J$ is a $J$ times smoothed version of the original data $D$,
and $w_j$ represents the structure of $D$ at scale $2^{j}$.  The
wavelet decomposition output consists of $J$ three-dimensional mother
fields $D_j$ and wavelets $w_j$ of size $n\times n \times
n$. Following the traditional indexing convention, we mark the mother
fields and wavelets of the finest scale with the index $j=1$. The
smoothed version of the original data, $D_J = D_0$, is the density
field found with the kernel of the scale, equal to the cell size of the
simulation $L/N_{\mathrm{grid}}$.

The wavelets can be found in a recursive manner, but we also needed to
evaluate the partial density fields (mother fields of different order)
for our analysis. Thus, we found the mother fields $D_j$ first by
convolving the field $D_{j-1}$ by the $B_3$ kernel of twice the scale
used for calculating the field $D_{j-1}$.  We then found the wavelets
of index $j$ by subtracting the mother density fields
\begin{equation}
w_j = D_{j-1} - D_j.
\end{equation}
In this construction, a wavelet of index $j$ describes density waves
between the scales $\Delta_{j-1} =l_c \times 2^{j-1/2}$ and $\Delta_{j} =
l_c\times 2^{j+1/2}$. The scales are the diameters of kernels used in calculating of
the density fields $D_{j-1}$ and $D_j$.

\end{appendix}

\end{document}